# Polarity control of carrier injection at ferroelectric/metal interfaces for electrically switchable diode and photovoltaic effects

D. Lee,[1] S. H. Baek,[2] T. H. Kim,[1] J.-G. Yoon,[3] C. M. Folkman,[2] C. B. Eom,[2,*] and T. W. Noh[1,†]

[1]ReCFI, Department of Physics and Astronomy, Seoul National University, Seoul 151-747, Republic of Korea

[2]Department of Materials Science and Engineering, University of Wisconsin-Madison, Madison, Wisconsin 53706, USA

[3]Department of Physics, University of Suwon, Suwon, Gyunggi-do 445-743, Republic of Korea

We investigated a switchable ferroelectric diode effect and its physical mechanism in Pt/BiFeO$_3$/SrRuO$_3$ thin-film capacitors. Our results of electrical measurements support that, near the Pt/BiFeO$_3$ interface of as-grown samples, a defective layer (possibly, an oxygen-vacancy-rich layer) becomes formed and disturbs carrier injection. We therefore used an electrical training process to obtain ferroelectric control of the diode polarity where, by changing the polarization direction using an external bias, we could switch the transport characteristics between forward and reverse diodes. Our system is characterized with a rectangular polarization hysteresis loop, with which we confirmed that the diode polarity switching occurred at the ferroelectric coercive voltage. Moreover, we observed a simultaneous switching of the diode polarity and the associated photovoltaic response dependent on the ferroelectric domain configurations. Our detailed study suggests that the polarization charge can affect the Schottky barrier at the ferroelectric/metal interfaces, resulting in a modulation of the interfacial carrier injection. The amount of polarization-modulated carrier injection can affect the transition voltage value at which a space-charge-limited bulk current–voltage ($J$–$V$) behavior is changed from Ohmic (*i.e.*, $J \propto V$) to nonlinear (*i.e.*, $J \propto V^n$ with n ≥ 2). This combination of bulk conduction and polarization-modulated carrier injection explains the detailed physical mechanism underlying the switchable diode effect in ferroelectric capacitors.

**I. INTRODUCTION**

Ferroelectric materials with their electrically switchable spontaneous polarization[1,2] can be used to control numerous functions by applying an external electric field. For example, polarization switching can be used to tune the charge transport properties at a ferroelectric/metal interface.[3–14] By choosing a metal with an appropriate work function, a Schottky barrier can be formed at the interface of various devices such as diodes. Then, by changing the polarization direction, the carrier injection through the barrier can be controlled, and, for example, the polarity of the diodes can be switched by simply applying an electric field.

Due to such intriguing possibilities, there have been numerous studies on adopting ferroelectric materials as electronic components, such as the aforementioned diodes.[3–7] Two decades ago, Blom *et al.*[4] reported a pioneering study on the diode effect in ferroelectric $PbTiO_3$ thin films. However, the diode effect was observed for only one polarity, with an Ohmic response observed for the other. Many similar experimental studies using other ferroelectric thin films followed, but most also reported the diode effect for only one polarity.[5–7] However, Choi *et al.*[8] recently reported a *switchable* diode effect in $BiFeO_3$ bulk single crystals, attributing this intriguing phenomenon to bulk effects. This switchable ferroelectric diode effect may enable novel applications, such as ferroelectric resistive memory with a non-destructive readout.[4,10,11] Additionally, Jiang *et al.*[15] reported switchable diode effects using $BiFeO_3$ thin-film capacitors, attributing the phenomena to oxygen-vacancy ($V_O$) accumulation near the electrode. However, more recently, some authors of Ref. 15 attributed the effects to polarization charge.[16] In these experiments, they used very leaky capacitors, so it is quite difficult to conclusively assign the observed changes in diode polarity to ferroelectric switching. In spite of these recent efforts,[8,15,16] the origin of the switchable diode effect is still not fully understood, with little consensus as to whether it is of bulk or interfacial origin.

In this paper, we investigate the electrically switchable diode and photovoltaic effects using high-quality Pt/BiFeO$_3$/SrRuO$_3$ thin-film capacitors. We demonstrate that the capacitor exhibits both forward and reverse diode effects, depending on the polarization direction within the ferroelectric layer. By comparing the changes in the transport characteristics with the polarization hysteresis loop, we show that the observed switchability between forward and reverse diodes is related to ferroelectric switching. In addition, we observe that the photovoltaic response can also be reversed by changing the polarization direction. These results demonstrate that both the diode and photovoltaic responses originate from band modification near the BiFeO$_3$/metal interfaces due to polarization charges. To explain these intriguing phenomena, we develop a thorough model of how the polarity of the diode effect can be controlled by the polarization direction. This understanding of the physical mechanism will provide an active path to the advanced design of switchable electronic thin-film devices.[17]

## II. EXPERIMENTS AND RESULTS

### A. Sample preparation

We used high-quality Pt/BiFeO$_3$/SrRuO$_3$ thin-film capacitors with a BiFeO$_3$ thickness of 400 nm. The films were fabricated with epitaxial BiFeO$_3$ sandwiched between a Pt top electrode and a single-crystal SrRuO$_3$ bottom electrode by sputtering onto SrTiO$_3$ (001) single-crystal substrates with a 4° miscut toward the [100] direction. The bottom electrode consisted of a 100-nm-thick epitaxial SrRuO$_3$ layer deposited on a SrTiO$_3$ substrate by 90° off-axis sputtering at 550°C.[18] We used a mixture of Ar and O$_2$ gases at a 3:2 ratio, maintaining the total pressure at 200 mTorr during the deposition. A 400-nm-thick epitaxial BiFeO$_3$ thin film was then grown on top of the SrRuO$_3$ bottom electrode by fast-rate off-axis sputtering at 690°C. The BiFeO$_3$ sputtering target contained an excess 5% Bi$_2$O$_3$ to compensate for the volatile Bi element. Here, we used a mixture of Ar and O$_2$ gases at a 3:1 ratio with the total pressure maintained at 400 mTorr. For the top electrode, a 40-nm-thick Pt layer was deposited at room temperature by on-axis sputtering. After the deposition, the Pt layer was photolithographically patterned to obtain the BiFeO$_3$ capacitors. The Pt top electrodes were between 50 and 200 μm in diameter. Figure 1(a) shows the x-ray diffraction (XRD) $\theta$–$2\theta$ scans of the BiFeO$_3$ films. Only the (001) and (002) pseudocubic reflections could be seen along with the substrate peaks, indicating the formation of a single-crystalline BiFeO$_3$ phase. Also, the inset of Figure 1(a) shows the surface morphology, measured by the atomic force microscopy (AFM) on the BiFeO$_3$ surface. The AFM image indicates that the film surface is smooth and has a morphology of the step-flow growth mode.[19]

**B. Electrically switchable diode effects**

*1. Electrical measurements*

The ferroelectric polarization–voltage (*P*–*V*) hysteresis loops were measured using a TF analyzer 2000 (aixACCT) at room temperature. As shown in Figure 1(b), our as-grown BiFeO$_3$ thin films had a large, approximately 65 μC cm$^{-2}$, remnant polarization along the [001] pseudo-cubic direction, which is consistent with earlier reports.[19–22] The coercive voltages of the films were about +4.5 V for positive bias and −5.5 V for negative bias. These differences in coercive voltages, called imprint, have been commonly observed in most BiFeO$_3$ thin films. Note that the *P*–*V* hysteresis loop was nearly rectangular, indicating no polarization relaxation, and very low leakage current through the sample. As discussed in Section III A, such a rectangular *P*–*V* hysteresis loop makes it easier to identify a process as directly linked to the ferroelectric polarization switching. We imaged ferroelectric domain patterns in our films by using a piezoelectric force microscope (PFM; XE-100, Park Systems). We measured the piezoelectric signal on the surface of Pt top electrode.

Our principle interest was in how the polarization direction inside the BiFeO$_3$ thin films affects the electrical transport properties. We measured the current density–voltage ($J$–$V$) at room temperature using a low-noise probe station and a picoampere meter (Keithley 236 source-measure unit). To obtain the downward (upward) polarization state, we applied a +10V (−10 V) external poling voltage to the capacitor. For the $J$–$V$ measurements, we used an applied bias between −0.8 and +0.8 V, much smaller than the coercive voltages, to ensure that no polarization switching occurred during the measurements. All of these electrical measurements were taken in the dark.

## 2. One-side diode effect in as-grown BiFeO$_3$ capacitors

All of our as-grown Pt/BiFeO$_3$/SrRuO$_3$ thin-film capacitors exhibited a one-side diode effect. As shown in Figure 2(a), with the negative poling (*i.e.*, upward polarization), the $J$–$V$ curve clearly indicated a diode effect with negative forward bias, called a reverse diode. (Similarly, for a positive forward bias, the diode effect was labeled a forward diode.) Conversely, with positive poling (*i.e.*, downward polarization), the $J$–$V$ curve did not show any diode effect between −0.8 and +0.8 V. Similar one-side diode effects have been reported with numerous ferroelectric thin-film capacitors,[4–7] but there has been little effort to understand their origin. In addition, such a one-side diode effect will not allow electrical switching between reverse and forward diodes.

This lack of switchability can be related to the formation of an interfacial defective layer at the Pt/BiFeO$_3$ top interface. All of our as-grown BiFeO$_3$ thin films on SrRuO$_3$ bottom electrodes were self-poled downward so that the negative polarization charge was built near the top BiFeO$_3$ surface.[20–22] During the high-temperature film deposition, as schematically shown in Figure 2(b), the downward self-polarization will cause the positively charged $V_O$ to migrate towards the top surface to compensate for the negative polarization charge.[23] Therefore, the $V_O$ migration can form a $V_O$-rich defective layer, which will remain between the Pt and BiFeO$_3$ layers in the capacitor. If the defective layer exists, it will seriously disturb the electrical processes at the Pt/BiFeO$_3$ top interface, especially the carrier injection under a positive bias. The detailed mechanism of how such a defective layer affects the $J$–$V$ characteristics will be discussed in Section III F.

We preformed thermal annealing experiments to confirm that the non-switchability between reverse and forward diodes (*i.e.*, the one-side diode effect) originated from an interfacial $V_O$-rich defective layer at the Pt/BiFeO$_3$ top interface. Figure 3(a) shows the $J$–$V$ characteristics of as-grown BiFeO$_3$ films within a wider external bias range. The $J$ value could be changed by about 2 orders of magnitude near the negative coercive voltage. However, as shown in the inset, a $V_O$-rich defective layer between the BiFeO$_3$ film and the Pt top electrode would not allow a similar change in $J$ under positive bias. To move the $V_O$-rich defective layer to the BiFeO$_3$/SrRuO$_3$ bottom interface, we switched the polarization of the film to upward and then annealed the BiFeO$_3$ sample at 200°C for 15 min in air. By changing the polarization direction, negative polarization charge was built in the bottom interface. Under high temperature, $V_O$ became highly mobile, and the $V_O$, originally located near the Pt/BiFeO$_3$ top interface, could migrate toward the BiFeO$_3$/SrRuO$_3$ bottom interface to compensate for the negative polarization charge. A $V_O$-rich defective layer could then become formed at the bottom interface, as shown in the inset of Figure 3(b). The $J$–$V$ curves in Figure 3(b) show that a change of $J$ value could now occur at the positive bias, consistent with the defect-layer configuration in the inset. This supports that the BiFeO$_3$ film transport properties can be strongly affected by the formation of a defective interfacial layer.

### *3. Electrical training to realize an electrically switchable diode*

To realize an electrically switchable diode, both BiFeO$_3$/metal top and bottom interfaces should be prepared such that the functionality of the device will not be hampered by the defective layers. Most of the $V_O$ formed during the fabrication process near the Pt/BiFeO$_3$ interface of the as-grown samples should be moved away from the top interface. To prepare such a $V_O$ sparse top interface, we applied a positive dc bias of +15 V to the as-grown sample for 30 min at room temperature. As schematically shown in Figure 3(c), this electrical-training process was expected to repel the $V_O$ near the top interface. During the electrical training, the compliance current was set at 1 µA to avoid electrical breakdown. (Note that the electrical training process did not appreciably affect the *P–V* hysteresis loop, as shown in Figure 1(b).) We chose a proper training time to ensure that most of the repelled $V_O$ would not move to the bottom interface and form other defective layers.

The transport characteristics between forward and reverse diodes of the electrically trained BiFeO$_3$ thin-film capacitors could be switched by changing the polarization direction of the ferroelectric layer. As shown in Figure 4(a), with the upward polarization, the BiFeO$_3$ capacitor exhibited the *J–V* curve of a reverse diode. Conversely, as shown in Figure 4(c), with downward polarization, it exhibited the *J–V* curve of a forward diode. These two curves demonstrate that electrically switchable diodes can be realized using ferroelectric thin-film capacitors.

**C. Electrically switchable photovoltaic effect**

We measured the photovoltaic response to be sure that the changes in the *J–V* curves were related to those of diode polarity and associated electronic structural changes. In a diode, the built-in internal field allows electric current to flow one way and prevents it from flowing the other. With illumination, the built-in field can separate the photoexcited electron-hole pairs. These separated electrons and holes move to opposite electrodes, resulting in a photovoltaic response. Thus, most diodes should exhibit a photovoltaic response.

We used white light, guided from a tungsten-halogen lamp onto the BiFeO$_3$ sample with an optical fiber, to make the photovoltaic measurements. Note that BiFeO$_3$ has a relatively small band gap of about 2.5 eV,[8] which corresponds to green light. Therefore, the BiFeO$_3$ layer can absorb a significant number of photons even with white light, leading to a photovoltaic response. A power density of the illuminated white light was less than 10 mW

cm$^{-2}$. The transmission of the 40-nm-thick Pt top electrode was estimated to be about 10–15 % for visible light.

Our trained BiFeO$_3$ thin-film capacitors showed electrically switchable photovoltaic response. As shown in Figures 4(b) and 4(d), they exhibited little $J$ without illumination. However, clear photovoltaic responses (*e.g.*, photocurrent at zero bias) occurred with illumination. In addition, the photovoltaic response could also be altered by changing the polarization direction. With the upward polarization, the photocurrent density was +0.18 μA cm$^{-2}$ and the photovoltage was −0.35 V, as shown in Figure 4(b); for the downward polarization, the photovoltaic effect was reversed and the photocurrent density was −0.10 μA cm$^{-2}$ and the photovoltage +0.25 V, as shown in Figure 4(d). This simultaneous observation of the switchable diode and photovoltaic effects in a single ferroelectric thin-film capacitor suggests that the ferroelectric, electronic, and optical properties are closely coupled.

## III. DISCUSSION

### A. Correlation between diode polarity and ferroelectric switching

The experiments showed that the switching between the forward and reverse diodes occurs very close to the ferroelectric coercive voltages. As schematically displayed in Figure 5(a), we initially poled the sample with a +10-V pulse and then applied an opposite polarity pulse ($V_p$) for 5 μs. We next measured $J$–$V$ in the low-bias region between −1 V and +1 V. As shown in Figure 5(b), the resultant $J$–$V$ curve initially indicated reverse-diode behavior that remained as long as $V_p \leq -5.0$ V. However, the curve changed suddenly to that of a forward diode with the application of $V_p = -5.5$ V. We executed similar measurements for the opposite, negative poling (Figure 5(c)). In this case, the diode polarity of the $J$–$V$ curves changed suddenly after applying $V_p = +4.5$ V (Figure 5(d)). Note that these $V_p$ values correspond to the coercive voltages in Figure 1(b).

To confirm that the switching between the forward and reverse diodes originated from the ferroelectric switching, we overlapped the response of the *J*–*V* curves with the *P*–*V* hysteresis loop, as shown in Figure 5(e). The *J* values at −1 V, as a function of $V_p$ in Figures 5(b) and 5(d), are plotted as red open circles and blue closed circles in Figure 5(e), respectively. This figure shows that the hysteresis behavior of the diode-polarity switching coincides with the ferroelectric-switching hysteresis. This confirms that the diode-polarity switching is closely coupled to the polarization direction of the ferroelectric layer.

As previously cited, and shown in Figure 5(e), our high-quality $BiFeO_3$ thin-film capacitors had a highly rectangular polarization hysteresis loop. For a ferroelectric thin-film capacitor with a slanted hysteresis loop, transient displacement current will occur due to the partial back-switching of polarization domains under a low applied bias. Such transient current makes it difficult to isolate the intrinsic conducting response in the *J*–*V* measurements. However, the rectangular polarization hysteresis loops in Figure 5(e) indicate that the measured *J*–*V* response should come from the intrinsic conduction through the $BiFeO_3$ capacitor, not from a transient displacement current. Additionally, some researchers recently reported electrically switchable diode polarity in highly leaky $BiFeO_3$ thin films, and attributed it to resistance changes due to $V_O$ movement.[15,24] If our observed changes in diode polarity had originated from the $V_O$ movement, the associated resistance changes would have been expected to be gradual. However, the sharp changes in *J* values near the coercive voltages in Figure 5(e) eliminate that possibility. Changes in diode polarity have an intrinsic origin that is closely related to those of the ferroelectric polarization direction.

## B. Conduction mechanism inside the $BiFeO_3$ film

Before addressing the detailed physical mechanism of the switchable ferroelectric diode, we investigated the dominant conduction route for carriers inside the $BiFeO_3$ films. The possible conduction mechanisms include interface-limited Schottky emission, space-charge-limited bulk conduction (SCLC), and bulk-limited Poole–Frenkel emission (PF).[25,26] According to the Schottky emission model, $J$ can be expressed as:

$$J \propto T^2 \exp\left[\frac{1}{k_B T}\left(\frac{q^3}{4\pi\varepsilon_0 K}E\right)^{0.5}\right], \quad (1)$$

where $q$ is the charge of the carriers, $K$ is the dielectric constant, and $T$ is the temperature. Thus, in Schottky emission, the data can be examined from the linear behavior of a $\ln(J)$ vs. $E^{0.5}$ plot. However, as shown in Figure 6(a), such a plot did not show linear behavior, particularly for positive bias. In addition, an attempt to obtain the value of $K$ by linearly fitting a limited region of 40 (V/cm)$^{0.5}$ < $E^{0.5}$ < 120 (V/cm)$^{0.5}$ yielded $K$ < 0.1, which is much too small when compared to the reported value of $K \sim 6.5$.[27] Moving to the PF emission model,

$$\frac{J}{E} \propto \exp\left[\frac{1}{k_B T}\left(\frac{q^3}{\pi\varepsilon_0 K}E\right)^{0.5}\right], \quad (2)$$

where the PF emission can be examined from the linear behavior in a $\ln(J/E)$ vs. $E^{0.5}$ plot. As shown in Figure 6(b), we also could not obtain a reasonable linear fit to the behavior for the PF emission model. Furthermore, the similarly obtained $K$ value was also below 0.1, still quite different from the reported value. Therefore, we excluded both the Schottky and PF emission models as dominant conduction mechanisms for the carriers inside the BiFeO$_3$ film.

Figure 7(a) schematically shows the predictions of the SCLC model.[25,26] At low $V$, the $J$–$V$ curve initially exhibits linear Ohmic behavior (*i.e.*, $J \propto V$), indicating that carriers are injected from the electrode and partially fill the shallow traps. With increasing $V$, the shallow traps become completely filled at the transition voltage $V_{tran}$, and then the partially filled deep traps start to play an important role in conduction ($J \propto V^n$ with n > 2). At even higher voltages, the deep traps are nearly filled, and deep-trap-free conduction occurs ($J \propto V^2$). Figure 7(b) shows the $\log(J)$ vs. $\log(E)$ curves with upward polarization inside our BiFeO$_3$ thin film. A power-law dependence yields a good fit, which is consistent with the predictions of the SCLC mechanism.[32] Thus, a detailed examination of *J–V* suggests that bulk-limited SCLC was a dominant conduction route for carriers inside BiFeO$_3$ films.

**C. Control of carrier injection by polarization change and its effect on SCLC**

The observed electrically switchable diode effect can be explained in terms of the polarization-direction dependence of the electrical characteristics at the BiFeO$_3$/metal interface. It is thought that the detailed BiFeO$_3$ and metal Fermi levels cause the BiFeO$_3$/metal interface to form a Schottky-like blocking barrier. Neglecting the polarization charge at the interface, the barrier height $\Phi$ is expected to be ~ 0.6 eV, depending on the impurity density in the films.[6,28] However, the large polarization charge can induce a significant change in band structure near the BiFeO$_3$/metal interface. According to a simple dielectric gap model,[13,29,30] this band modification due to the polarization charge is accompanied with a change in the barrier height: $\Delta\Phi = P\delta/\varepsilon_0\varepsilon_{st} = \pm 0.6$ eV, with $P = \pm 65$ μC cm$^{-2}$, $\delta = 1$ nm, and $\varepsilon_{st} = 100$. Thus, for the upward polarization state, the Pt/BiFeO$_3$ and BiFeO$_3$/SrRuO$_3$ interfaces can have blocking and nearly non-blocking contacts, respectively, as shown in Figure 8(a). Conversely, for the downward polarization state, the Pt/BiFeO$_3$ and BiFeO$_3$/SrRuO$_3$ interfaces can have nearly non-blocking and blocking contacts, respectively. Thus, the electrical characteristics at both interfaces can be controlled simply by changing the polarization direction of the ferroelectric layer.

Also according to SCLC theory,[25,26] transition from the Ohmic to the nonlinear regions occurs at

$$V_{tran} \sim \frac{q(H_a - p_t)d^2}{2K}, \qquad (3)$$

where $H_a$ is the total trap density, $p_t$ is the density of trapped carriers, and $d$ is the sample thickness. As the carriers inside the film increase by carrier injection from the electrode, $p_t$ will increase effectively, resulting in the decrease of $V_{tran}$. Thus, if the electronic contact at the interface can be modulated between blocking and nearly non-blocking contacts, the $V_{tran}$ value can be controlled. For a nearly non-blocking contact, a high injection of carriers can induce a low $V_{tran}$, so the transition from the Ohmic to nonlinear regions can easily occur. On the other hand, for blocking contacts, $V_{tran}$ will be quite large, and the transition is not likely to occur during our J–V measurements. Therefore, according to the SCLC theory, the interface contact type and associated carrier injection from electrode should play an important role to determine the electrical properties of the Pt/BiFeO$_3$/SrRuO$_3$ capacitors.

**D. Explanation of the electrically switchable diode**

Let us consider the situation with the upward polarization state inside the BiFeO$_3$ films.[31] As already shown in Figure 7(b), the J–V curves are consistent with the SCLC mechanism predictions. Most BiFeO$_3$ films exhibit p-type conduction as a result of Bi loss, where the holes become the dominant carriers.[5,6] For a positive applied bias, the injection of hole carriers at the blocking Pt/BiFeO$_3$ contact becomes suppressed, as schematically shown in Figure 8(b). Thus, the J–V curve initially exhibits a linear behavior. At higher positive V, the blocking contact dominates, and the strong carrier injection suppression results in a sublinear region,[25,26] which is indicated by the green squares in Figure 7(b). Conversely, for a negative applied voltage, there is high hole injection from the nearly non-blocking BiFeO$_3$/SrRuO$_3$ contact, as indicated in Figures 8(c) and 8(d). This high hole injection causes a transition from the initial linear J–V curve to a rapidly increasing nonlinear J–V curve at $V_{tran} = -0.15$ V, as indicated by the blue and red squares in Figure 7(b).

Based on this J–V curve, we can explain the reverse diode behavior in Figure 4(a). When we increase V from negative to positive, we are initially following the dashed arrow and then the dotted arrow. The corresponding changes of J are marked with the same kinds of arrows in Figure 4(a). Therefore, the reverse diode behavior can be explained in terms of carrier injection changes in the upward polarization state. Similarly, for the downward polarization state, the forward diode in Figure 4(c) can be explained by the reversing of the nearly non-blocking and blocking contacts at the BiFeO$_3$/metal interfaces. Therefore, the switchability between the reverse and forward diode effect, observed in Figure 4, originates from the band modification and associated changes in the electric characteristics of the BiFeO$_3$/metal interfaces that depend on the polarization direction.[32]

**E. Explanation of the electrically switchable photovoltaic effect**

The polarization direction-dependent band modification of the BiFeO$_3$ capacitor can also explain the switchable photovoltaic effects presented in Figures 4(b) and 4(d). It is well known that the built-in field caused by the electronic band-bending can generate a photovoltaic effect.[25] Figures 9(a) and 9(b) show the built-in field ($E_{bi}$) and the operational principle of the photovoltaic effect for the upward and downward polarization states, respectively. The downward $E_{bi}$ for the upward polarization generates a positive photocurrent and a negative photovoltage in Figure 4(b). The $E_{bi}$ direction, due to band bending, can be reversed by polarization switching in our BiFeO$_3$ capacitor. The upward $E_{bi}$ for the downward polarization state results in a negative photocurrent and a positive photovoltage in Figure 4(d). These explanations of both the diode and photovoltaic responses based on a single physical picture also support that a simple electrical switching of the ferroelectric polarization direction can be used to tune the polarity of both devices.

**F. Explanations on annealing effects**

In Section III C and D, we showed that the polarization charge could affect interfacial band structures and result in a switchable diode in our BiFeO$_3$ capacitors. Specifically, we suggested that the Schottky barrier and associated band-bending character of the BiFeO$_3$/metal interfaces could be changed by a reversal of polarization direction. Such a change results in variations in the carriers injected from an electrode into the BiFeO$_3$ layer. In the case of an ideal BiFeO$_3$ capacitor, the hole carriers can be injected from the top (bottom) electrodes when a positive (negative) bias is applied. However, if a defective layer is located at the interface (*i.e.*, between the BiFeO$_3$ and metal layers), the carrier injection process could be significantly disturbed.

In Figures 2 and 3(a), it was assumed that the defective layer was located at the Pt/BiFeO$_3$ top interface where the hole injection from the BiFeO$_3$/SrRuO$_3$ bottom interface would not be affected by the defective layer. With negative bias, the *J* value can vary significantly, depending on the polarization direction of the BiFeO$_3$ layer, so it can exhibit a large resistive change. However, with positive bias, the hole injection from the Pt/BiFeO$_3$ top interface will be significantly disturbed, so that it will not show an observable resistive change. However, in Figure 3(b), where it was assumed that the defective layer is located at the BiFeO$_3$/SrRuO$_3$ bottom interface, the modulation due to the polarization direction change will only work for the positive bias side. Therefore, our model successfully explains the results of the thermal annealing shown in Figures 3(a) and 3(b).

**G. Tailoring of the diode and photovoltaic effects by ferroelectric domain manipulation**

One important possibility for electrically switchable diode and/or photovoltaic devices is that we can tailor their electrical and photovoltaic properties by adjusting the amount of upward and downward polarization inside the BiFeO$_3$ thin-film capacitor. To demonstrate, as shown in the inset of Figure 4(e), we generated a domain pattern with nearly equal-portioned upward and downward polarizations. (We made this nearly 50% domain pattern by adjusting the strength (i.e., height and width) of applied voltage pulse. For example, we applied the voltage pulse with a width of ~5 ms and a height of ~4.0 V to the capacitor with the up polarization state.) According to the aforementioned mechanism, the electrical and photovoltaic response of a multi-domain configuration should be interpreted as a parallel connection of forward and reverse diodes. Thus, with nearly the same amount of upward and downward domains, the *J–V* curve becomes nearly symmetric. Furthermore, the rectifying diode effect of the BiFeO$_3$ thin-film capacitor can be turned off, as shown in Figure 4(e).[33] In addition, the parallel connection of opposite diodes can also cancel out the photovoltaic effect, even under illumination, as shown in Figure 4(f).

By adjusting the proportion of upward and downward domains, we can generate devices with numerous stable states, which are labeled multilevels. As confirmation, we fabricated a three-state non-volatile photovoltaic device using polarization states with 100%, nearly 50%, and 0% upward domains. Figure 4(g) shows the ferroelectric polarization dependence of the photocurrent density at zero bias. We first measured the photocurrent for the upward polarization state for 100 cycles. We then reversed the polarization direction by applying a +10-V pulse and measured the photocurrent at zero bias. After 100 measurements, we created an upward/downward mixed-polarization state by applying a moderate voltage and again measured the photocurrent for 100 cycles. This procedure was repeated four times. As shown in Figure 4(g), the polarity and magnitude of the photocurrent could be controlled electrically by manipulation of the ferroelectric domain. This demonstrates that our ferroelectric thin-film capacitors can be used to control the photovoltaic effect in a reversible, non-volatile, highly tunable manner.

## IV. SUMMARY

We demonstrated polarization-direction control of the electrical characteristics at BiFeO$_3$/metal interfaces and achieved switchable ferroelectric thin-film diode and photovoltaic devices. We suggested that the polarization-modulated interfacial carrier injection could affect the bulk conduction characteristics, resulting in the switchable diode effect. This interfacial effect gives additional functionality to the bulk properties of BiFeO$_3$ thin films, enabling new ferroelectric/multiferroic applications. Our discovery provides an active path for the advanced design of novel multifunctional devices, such as switchable optoelectronic devices and resistive switching memory devices.

**ACKNOWLEDGMENTS**

D.L. and S.H.B. contributed equally to this work. This research was supported by the National Research Foundation of Korea, funded by the Korean Ministry of Education, Science, and Technology through Grants 2009-0080567 and 2010-0020416. Work conducted at the University of Wisconsin–Madison was supported by the Army Research Office through Grant W911NF-10-1-0362, the National Science Foundation through Grant ECCS-0708759, and a David & Lucile Packard Fellowship (C.B.E.). D.L. acknowledges support from the POSCO TJ Park Doctoral Foundation.

[31]Corresponding data for the downward polarization state are similar. Due to the page limits, they are not shown here.

[32]Pt and $SrRuO_3$ electrodes we used have slightly different work functions, whose values are 5.1~5.8 eV and ~5.2 eV, respectively. Thus, even without the polarization charge effect, the heights of Schottky barriers at $Pt/BiFeO_3$ and $BiFeO_3/SrRuO_3$ interfaces can be different from each other. This might result in rather different characteristics (e.g., the current level and on/off ratio) of the diode effects according to the polarization directions, as shown in Figures 4(a) and 4(c).

[33]Previous studies reported that there can occur a relatively high conduction at domain walls of $BiFeO_3$ (Ref. 12). The nearly 50% mixed domain configuration includes domain walls, and thus there can be an additional contribution from the domain walls to conductive behavior. However, since domains and domain walls can contribute to the conduction together in our capacitor geometry, we cannot distinguish the domain walls' effect from the domains' effect at this moment.

<Figure captions>

FIG. 1. (Color online) (a) XRD $\theta$–$2\theta$ scan of the BiFeO$_3$ film on SrRuO$_3$/SrTiO$_3$ substrate. The inset in (a) shows the AFM image of the BiFeO$_3$ film surface. (b) Rectangular polarization–voltage hysteresis loop of the as-grown BiFeO$_3$ capacitor (closed triangles) and the electrically trained capacitor (open squares).

FIG. 2. (Color online) (a) $J$–$V$ curves for the as-grown BiFeO$_3$ thin capacitor. (b) $V_O$-rich defective layer at the as-grown BiFeO$_3$/Pt interface resulting from downward self-polarization. Green, blue, red, and grey circles indicate Bi, Fe, O, and Pt, respectively.

FIG. 3. (Color online) $J$–$V$ characteristics of BiFeO$_3$ capacitors with (a) as-grown and (b) modified defect structures. The red layer in the schematic inset indicates the $V_O$-rich defective layer. (c) Applying a positive dc bias can recover the $V_O$-rich defective layer by repelling the $V_O$.

FIG. 4. (Color online) $J$–$V$ curves (open squares) for the electrically trained BiFeO$_3$ thin capacitor, measured in the dark for the (a) upward-, (c) downward-, and (e) upward/downward mixed-polarization domains. The upper-left insets show the ferroelectric domains measured by the out-of-plane piezoelectric atomic force microscope. $J$–$V$ curves (closed squares) for the electrically trained BiFeO$_3$ thin capacitor measured under illumination for the (b) upward-, (d) downward-, and (f) upward/downward mixed-polarization domains. (g) Cycling test of the three-state non-volatile photovoltaic effect.

FIG. 5. (Color online) Schematics of the $J$–$V$ measurement sequence after poling with (a) +10-V and (c) −10-V pulses. (b), (d) $J$–$V$ curves measured after applying each opposite-

polarity pulse with an incrementally increasing amplitude ($V_p$). (e) Plot of $J$ measured at −1 V as a function of $V_p$, overlapped onto the polarization–voltage hysteresis loop.

FIG. 6. (Color online) (a) ln($J$) vs. $E^{0.5}$ plots for the upward polarization state. (b) ln($J/E$) vs. $E^{0.5}$ plots for the upward polarization state.

FIG. 7. (Color online) (a) Typical $J$–$V$ curve for space-charge-limited conduction. (b) Experimental $J$–$V$ curve for the upward polarization state. We draw black solid lines to indicate power-law behaviors of $J$–$V$ curves.

FIG. 8. (Color online) (a) Band characteristic for the upward polarization state. Dashed black lines indicate the band characteristic when the effect of polarization is not considered. Solid lines indicate band bending modified by the polarization charge. Expected conductions for the upward polarization state when (b) positive bias is applied to the Pt top electrode. Expected conductions for the upward polarization state when negative bias with (c) $-V_{tran} < V < 0$ and (d) $V < -V_{tran}$ is applied to the Pt top electrode.

FIG. 9. (Color online) Schematic description of the photovoltaic effect for (a) the upward and (b) downward polarization states. The slope of the band edges at the interfaces can generate a built-in field ($E_{bi}$).

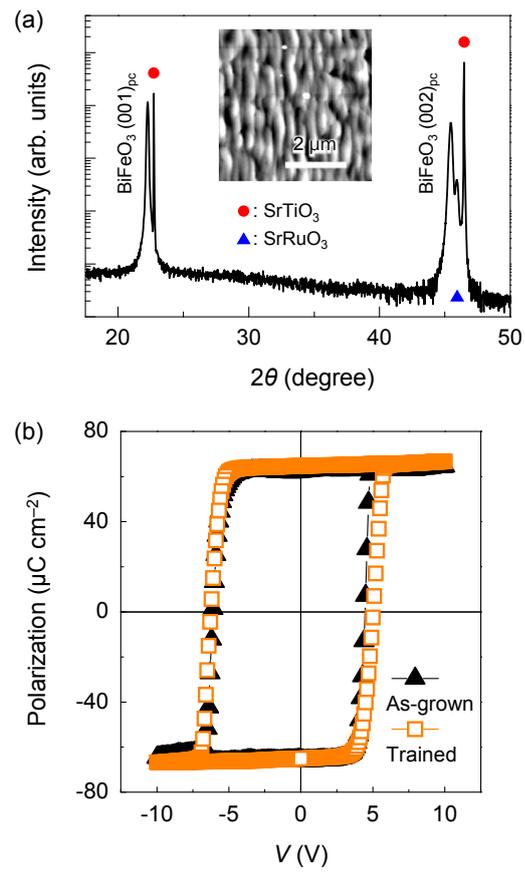

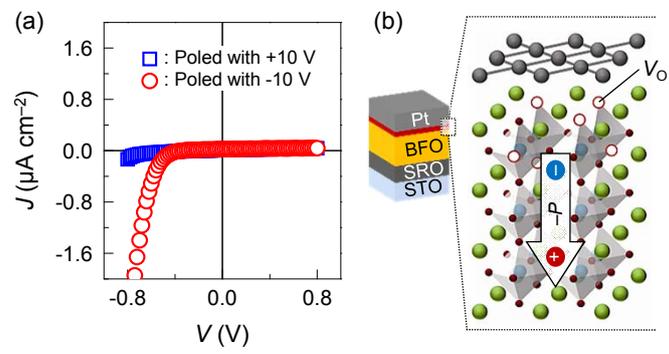

D. Lee *et al.*, Figure 3

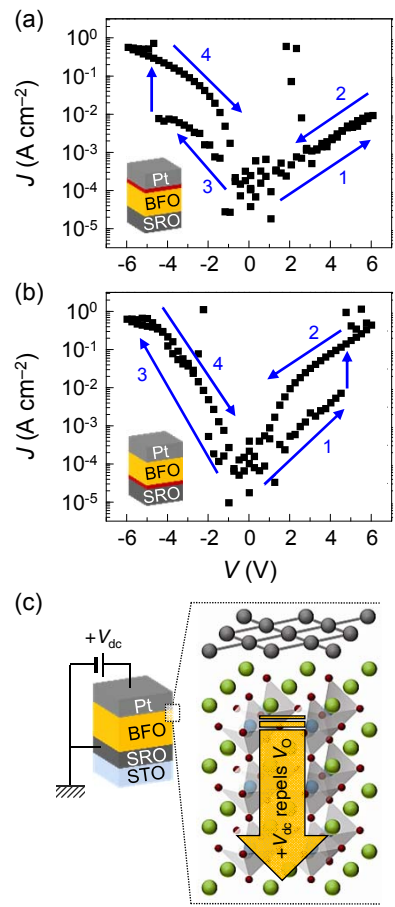

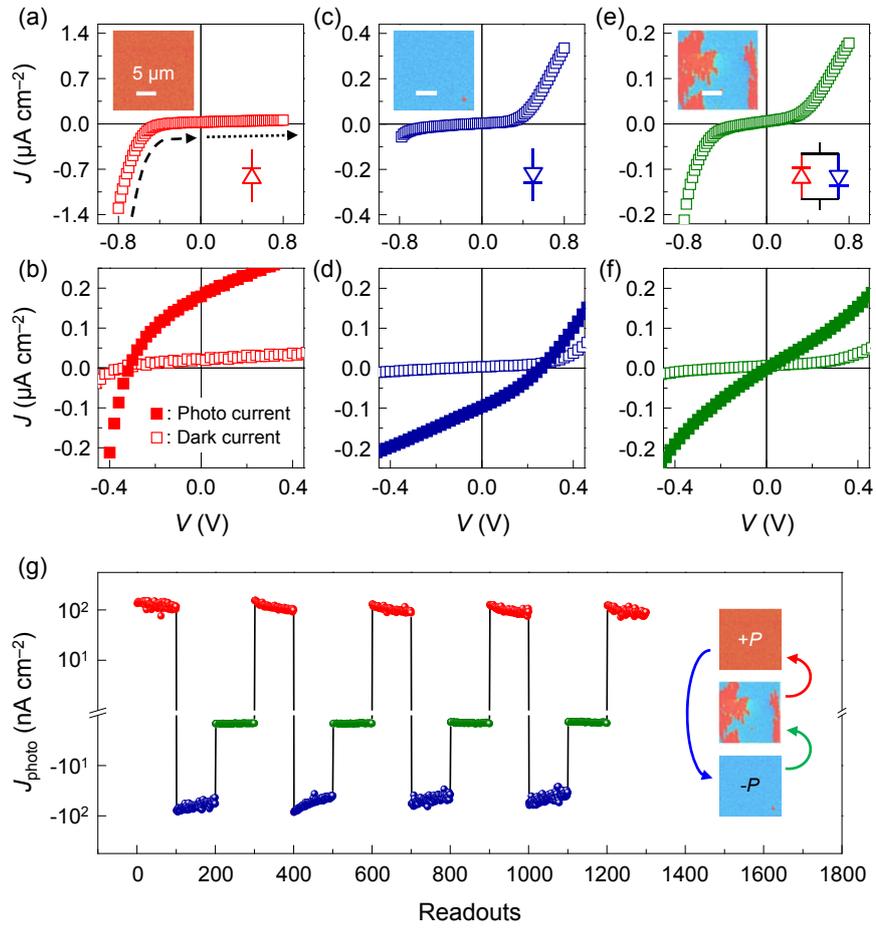

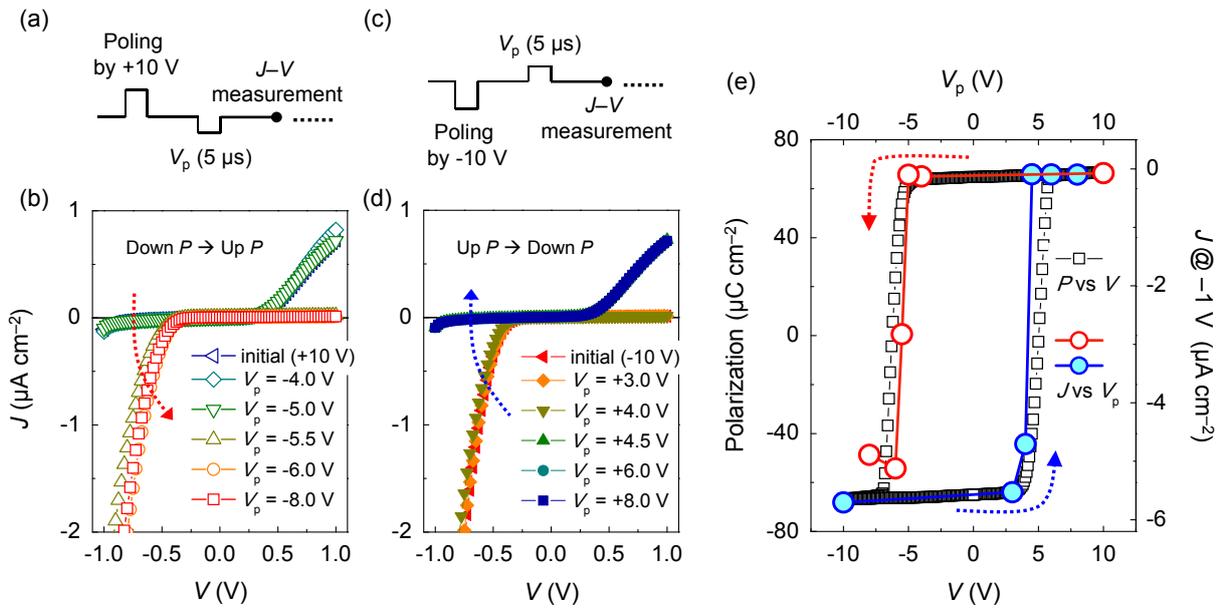

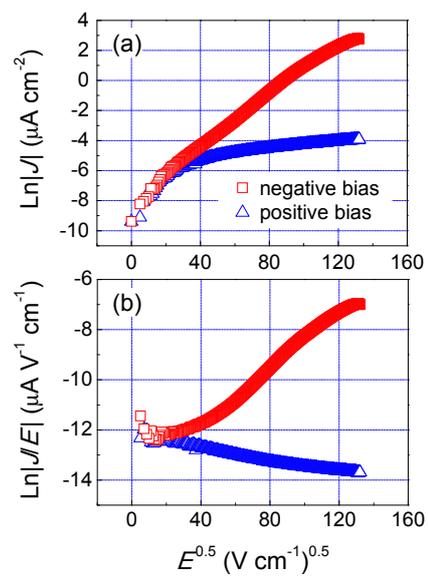

D. Lee *et al.*, Figure 7

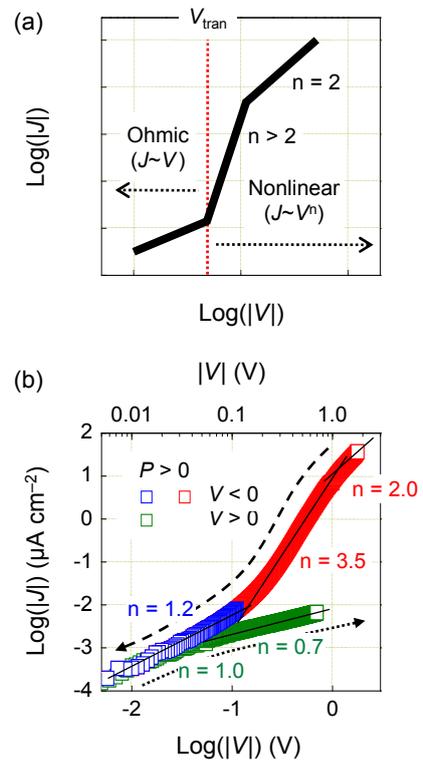

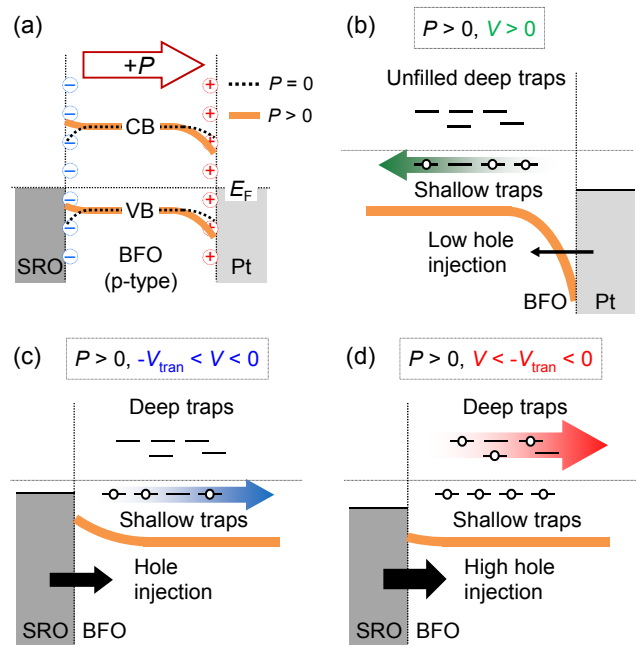

D. Lee *et al.*, Figure 9

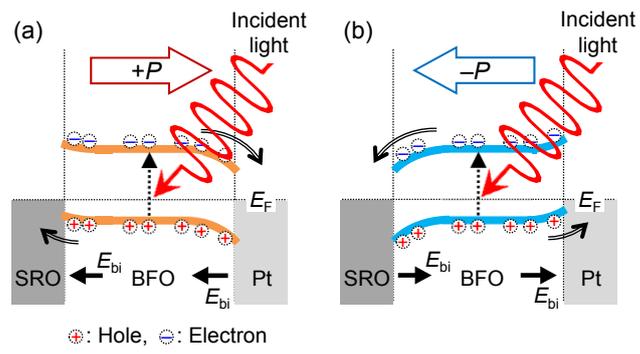